# A Microsimulation Analysis of the Distributional Impact over the Three Waves of the COVID-19 Crisis in Ireland[1][2]

Cathal O'Donoghue*, Denisa M. Sologon**, Iryna Kyzyma**, John McHale*

* National University of Ireland, Galway

** Luxembourg Institute of Socio-Economic Research

**Abstract**

This paper relies on a microsimulation framework to undertake an analysis of the distributional implications of the COVID-19 crisis over three waves. Given the lack of real-time survey data during the fast moving crisis, it applies a nowcasting methodology and real-time aggregate administrative data to calibrate an income survey and to simulate changes in the tax benefit system that attempted to mitigate the impacts of the crisis. Our analysis shows how crisis-induced income-support policy innovations combined with existing progressive elements of the tax-benefit system were effective in avoiding an increase in income inequality at all stages of waves 1-3 of the COVID-19 emergency in Ireland. There was, however, a decline in generosity over time as benefits became more targeted. On a methodological level, our paper makes a specific contribution in relation to the choice of welfare measure in assessing the impact of the COVID-19 crisis on inequality.

[1] The Authors are grateful to the Irish Health Research Board and Irish Research Council for funding of this research.
[2] Corresponding author: Denisa Sologon (address: LISER, 11 Porte des Sciences, L-4366 Esch-sur-Alzette, Luxembourg; email: Denisa.Sologon@liser.lu).

# 1. Introduction

"COVID-19 does not discriminate" is a phrase that has been used extensively. A google search reveals 80300 uses of the phrase. Users vary from the United Nations Human Rights, media outlets, and academic papers. However Patel et al. (2020) argue convincingly that this "is a dangerous myth, sidelining the increased vulnerability of those most socially and economically deprived." The COVID-19 crisis has brought a major impact on health, on economies and on personal incomes in a highly asymmetric way, affecting different people in different ways.

Given the rapid spread of the COVID-19 virus, governments have had to respond rapidly and quite severely to flatten the curve and slow the spread of the virus. This has had significant implications on many aspects of life, acting differentially on different groups. In many cases, interventions have been crude, by necessity, given the paucity of data and diagnostics necessary for more targeted policy and given the need for speed to stop the transmission of the virus. The impact of the COVID-19 virus on people is multi-faceted (O'Donoghue et al., 2020). In addition to the medical and health impacts, there are direct and indirect economic and social impacts, including market effects and policy mitigation.

Given the increase in home working, there are also important impacts on disposable income in relation to work related costs. Indeed, households who have not lost their income will indirectly have received a windfall gain in terms of an increase in purchasing power during the crisis as a result of lower work-related costs or higher pandemic benefits than previous work incomes. Lower opportunities to spend has meant an overall large increase in savings. Finally, there may be differential price impacts.

One of the challenges in trying to understand the distributional impact of a fast moving crisis is the time lag between the availability of micro data and the crisis. There is a growing literature in utilising nowcasting, taken from the macro literature (Giannone et al., 2008) to adjust survey data and make it consistent with current macro-economic trends.[3] However, improving policy design requires knowledge about policy incidence and change on the distribution that raises the need in disaggregated nowcasting for microsimulation purposes, reviewed in O'Donoghue and Loughrey (2014).

The majority of the microsimulation literature on nowcasting utilises the EUROMOD tax-benefit model. For example, Leventi et al. (2014) and Navicke et al. (2014) applied the nowcasting method to update poverty indicators (calibrating within population sub-groups) by linking income surveys such as the Survey of Income and Living Conditions to aggregated labour market conditions from the European Labour Force Survey. The approach has been applied during COVID-19 for a number of European Countries (Beirne et al. 2020; Brewer and Gardiner, 2020; Brewer and Tasseva, 2020; Bronka et al., 2020; Figari and Fiorio, 2020).

There have been a number of alternative approaches. Addabbo et al. (2016) improved the heterogeneity of transitions by estimating parametric equations to capture employment changes using EUROMOD instead of cell based Monte-Carlo simulations. Instead of calibrating income surveys to labour force surveys, Carta (2020) imputed labour income onto recent labour force data, albeit only looked at the distribution of market income. O'Donoghue et al. (2020) and Sologon et al. (2020) also utilised a parametric approach, but drawing upon the alignment methodology of the dynamic microsimulation literature described in Li and O'Donoghue (2014).

---
[3] Another literature has produced nowcasted aggregate poverty rates (Álvarez et al., 2014).

Li et al. (2020) took a different approach along the lines of the static ageing literature (Immervoll et al., 2005) – a semi-parametric perspective – drawing upon the methodology of DiNardo, Fortin and Lemieux (1996). Although incorporating flexible distributional forms, the semi- or non-parametric methods have a risk of relying on small cells sizes, particularly when there are a lot of dimensions used in the reweighting (Klevmarken, 1997). Therefore, in this paper, we utilise a parametric approach initially applied in O'Donoghue et al. (2020) in order to study the impact of the COVID-19 crisis on the distribution of household income. We extend this analysis from an initial impact assessment of the start of the crisis to an evaluation of changes over the course of the crisis.

In this paper, we take Ireland as a case study. It is a country that was significantly affected by the Great Recession (O'Donoghue et al. 2013, 2018). At the time of writing, Ireland has also been through three phases of the COVID-19 crisis (Figure 1). In response to this crisis, the State instituted three main support payments: the Pandemic Unemployment Payment (PUP) targeting those who were laid off work because the business had to close as a result of the pandemics; the COVID Enhanced Illness Benefit (CEIB) for those who were out of work either due to contracting the COVID-19 virus or because they had to self-isolate due to a close contact; and various COVID Wage Subsidies (CWS) aimed at supporting employers to maintain employment contract with employees despite the fall in revenues. Figure 2 highlights the growth in demand for these payments over the crisis, peaking at nearly a million recipients of all benefits for a work force of just over 2 million.[4]

**Figure 1. Daily number of COVID-19 cases in Ireland**

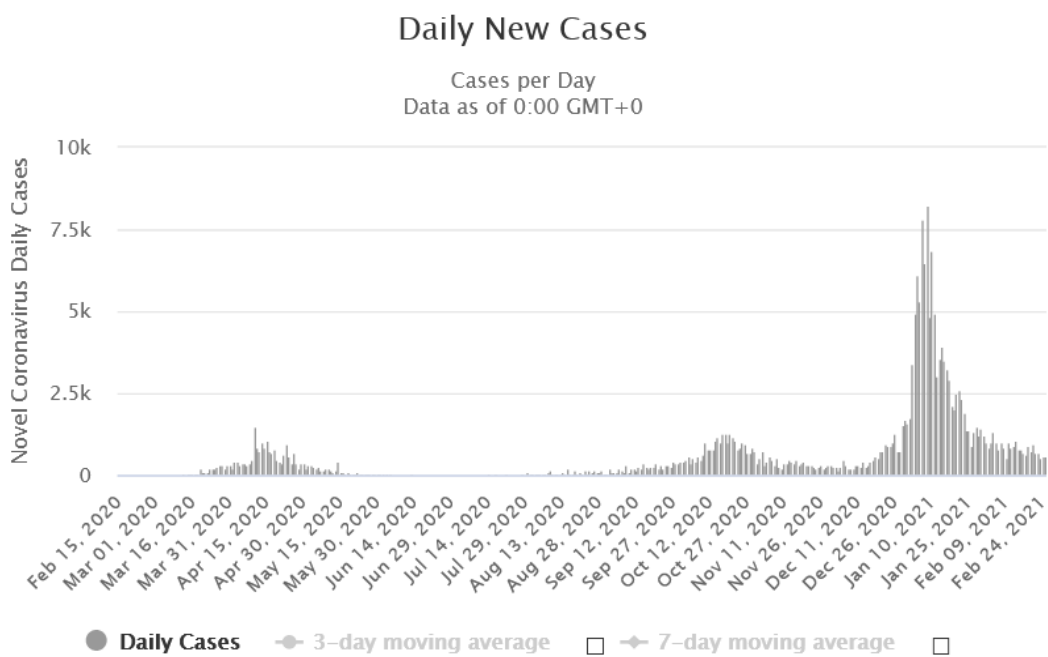

Source: Worldometer data.

---

[4] Figure A.1 in Appendix A also outlines the phases of economic restrictions over the course of the two waves, with the trend in expenditure per Revolut user over the period. It shows the decline and then recovery of expenditure in the first wave and subsequent fall again in the second wave as restrictions, particularly in retail and hospitality sectors, were applied.

**Figure 2. Trends in the recipients of the pandemic related payments**

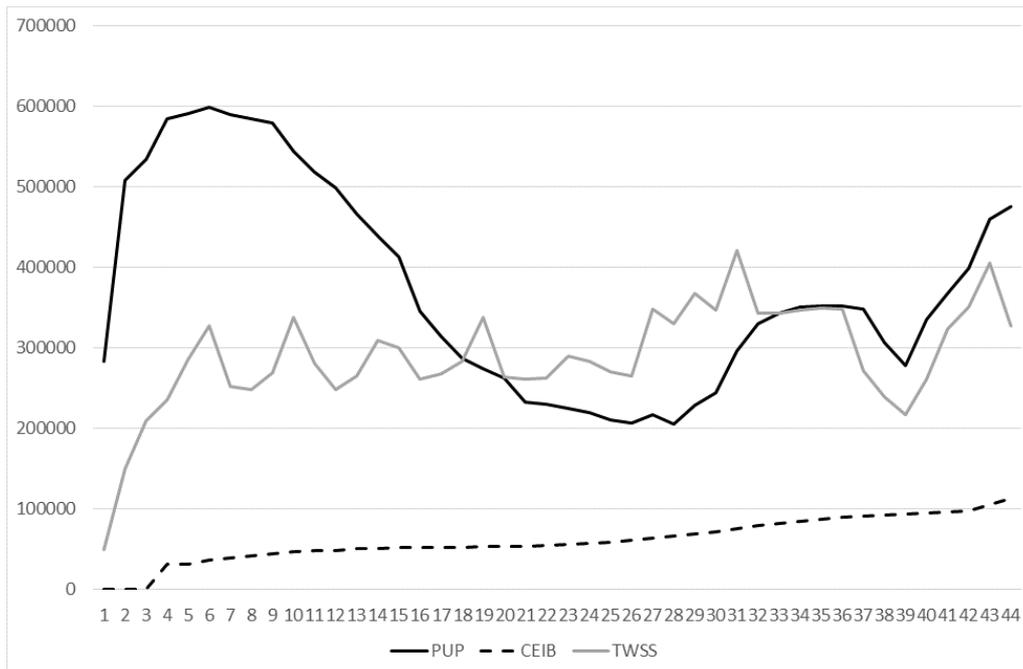

Note: CWS stands for various COVID Wage Subsidies; PUP stands for Pandemic Unemployment Payment; CEIB stands for COVID Enhanced Illness Benefit.

Because of the multi-faceted nature of the shock, affecting market incomes but also demands for childcare, commuting costs and mortgage costs, we utilise a non-standard definition of disposable income, which relies on adjusting household disposable equivalized income for work-related and housing expenditures, as well as for some capital losses. At the core of our nowcasting approach lies a household income generation model (Sologon et al., 2021). The model relies on data from the Survey of Income and Living Conditions (SILC) calibrated to account for the labour market and policy impacts of COVID-19 using administrative data and the Labour Force Survey (LFS). This allows generating counterfactual income distributions as a function of more timely external data than the underlying income survey.

## 2. Methodology

We model the distributional impact of COVID-19 and its associated policy responses. From an economic perspective, the COVID-19 virus affects people differentially across different dimensions, including:

- Those who get sick have a spectrum of consequences from self-isolation and time away from work, study and family to hospitalization and mortality.
- A far greater proportion of the population are affected by closing businesses and their loss of income or the social implications of cocooning. Unlike a typical demand shock, the biggest impacts are felt by those in so called non-essential businesses. The income implications are varied from total loss of income to increased income in some retail businesses.
- This impact on the economy has seen a large fall in capital asset values.

- Various policy responses such as the Pandemic Unemployment Payment or the Temporary Wage Subsidy will have mitigated some of the impact of job loss or wage reduction, but not fully.
- Agreements with banks in relation to mortgages, a freeze on evictions and supports for childcare providers will improve the cash flow of households.
- However, some households who have not lost their income will indirectly have received a windfall gain in terms of an increase in purchasing power during the crisis as a result of lower work-related costs or higher benefits than work income.

Ideally, in undertaking an analysis of these effects, we would use household survey data to assess distributional impacts. However, there is a time lag between collection and release for research and analysis. For example, the main survey used that contains the income situation of households is the Survey of Income and Living Conditions. The most recent analysis undertaken is for 2018. In normal times, a lot happens in a two-year period, but in a crisis the changes are so significant that such a lag can mean the data is relatively meaningless.

There are more recent datasets available that can assist such as the Labour Force Survey, which is available on a quarterly basis at a six-week lag or the Live Register data and Price data that is available on a monthly basis on a short lag. However, these datasets do not contain income information. In order to improve the targeting of policies with minimal cost we need to understand in real-time how the shock affects the incomes of different types of households.

*Nowcasting*

From a methodological point of view, the key challenge is the lack of up to date information. We propose to overcome this data gap by using a "nowcasting" methodology (O'Donoghue and Loughrey, 2014) with recent data on employment and prices to calibrate a microsimulation model of household incomes, taxes and benefits to produce a real-time picture of the population and who is affected differentially (O'Donoghue, 2014; Atkinson et al, 2002).

Our paper goes beyond existing methods that apply price inflation factors, change proportionally the employment rate in specific industries and apply tax-benefit transformations to explain the policy consequences (Navicke et al., 2014). In periods of significant volatility with large and rapid changes in the structure of the economy, such as during the COVID-19 crisis, it is more appropriate to utilise a dynamic-type income generation model approach to update the data and to capture the heterogeneity of changes in the population (see Li and O'Donoghue, 2014; Bourguignon et al., 2001). We follow a similar approach as studies looking at the first 2 waves of COVID-19 (e.g. O'Donoghue et al. (2020), Sologon et al. (2020)). At the core of our nowcasting approach, lies an income generation model, which consists of a system of equations that describe the household market income distribution as a function of personal and household attributes (Sologon et al. 2021). The parameters of the income generation model are used to simulate counterfactual distributions of market incomes under alternative scenarios[5]. In order to convert market incomes into disposable incomes, we utilize the NUI Galway tax-benefit simulator (O'Donoghue et al, 2013).

This methodology was already used to understand the drivers of cross-national difference in inequality in disposable income (Sologon et al. 2021) and the drivers of changes in disposable income inequality (Sologon et al., 2019). In this paper, we use the infrastructure to update the latest available survey data by calibrating the simulations to external macro controls to reflect more timely Live Register, Price and Labour Force Survey data to undertake the COVID distributional impact assessment.

---

[5] Please refer to Sologon et al. (2021) for an in-depth discussion of simulating counterfactuals.

To accommodate the nature of the shock and the multi-faceted impact on household living standards, our core welfare variable of interest is an augmented definition disposable income which accounts for work-related and housing costs. Following the standard definition, disposable income, $Y_{D,t}$, at time $t$ depends upon market income $Y_{M,t}$, benefits $B(Y_{M,t}, Z_t, \theta_t^B)$ and taxation $T(Y_{M,t}, Z_t, \theta_t^T)$, which are in turn dependent upon personal skills, family characteristics, Z, and tax-benefit parameters $\theta$. Our analysis adjusts disposable income for:
- work-related expenditures $C_t$:
- housing costs $H_t$:
- capital losses $Q_t$

$$Y_{D,t} = Y_{M,t} - T(Y_{M,t}, Z_t, \theta_t^T) + B(Y_{M,t}, Z_t, \theta_t^B) - H_t - Q_t - C_t.$$

To some extent this turns the clock back to microsimulation analyses from the 1980's where disposable income net of housing costs were used occasionally (Atkinson et al., 1993; Atkinson, 1995).

The nowcasting processes involves a number of components:
- Estimation and simulation of a system of hierarchically structured, multiple equations, known as an income generation model that describe the presence $(I_{i,t})$ and level $(Y_{i,t})$ of market incomes (Sologon et al. 2021)

$$Y_{M,t} = \sum_{i=1\ldots m} Y_{i,t}^* = \sum_{i=1\ldots m} \left\{ Y_{i,t}(Z_{i,t}^Y, \theta_{i,t}^Y, \varepsilon_{i,t}^Y) \times I_{i,t}(Z_{i,t}^I, \theta_{i,t}^I, \varepsilon_{i,t}^I) \right\}$$

- A tax-benefit model, described in O'Donoghue et al. (2013), to simulate tax-benefit changes T(), B()
- Income indexation – the change in the level of income resulting from changes to average wages $Y_{i,t}()$

The mechanism has at its core a generic household income-generation model (IGM) similar to Sologon et al. (2021). The labour market module estimates the statistical distribution of labour market factors: the probability to be at work, to earn income from salaried employment or self-employment, the occupational, sector and industry, choices, the probability of being unemployed, retired (if not working), the prevalence of income sources (investment income, property income, private pension, other income), the probability of paying for housing (home owner, mortgage, rent), the probability of paying contributions (private pensions), the probability of having child care.

The market composition module involves two estimation techniques: (i) binary models for binary outcomes, and (ii) multinomial models for m outcomes, $m > 2$. In order to use the estimated probabilities from logistic models within a Monte Carlo simulation, we draw a set of random numbers such that we predict the actual dependent variable in the base year (see Sologon et al. 2021 for the method). The disturbance terms are normally distributed, recovered directly from the data for those with observed incomes, or generated stochastically for those without a specific income source in the data.

At each step, we retrieve the parameters estimates and the individual specific errors for each estimated model, to be subsequently used in simulating counterfactuals. We use the IGM to simulate the impact of changing economic conditions over time. Bourguignon et al. (2002) and Sologon et al. (2019) used a similar methodology to disentangle the impact of macro-economic changes on inequality by generating counterfactual distribution - transformations of the income generation process by 'swapping coefficients' between years for the various transformations.

In nowcasting, the simulations involve calibrating econometrically estimated equations in the income generation model to external control totals made available in more timely data than the estimation data, only available at a lag. The calibration mechanism or alignment is drawn from the dynamic microsimulation literature (Li and O'Donoghue, 2014) and aims to calibrate the microsimulation model in order to match the simulated output to exogenous totals, particularly in relation to the structure of the labour market (Baekgaard, 2002). In our model we utilise three types of alignment for binary discrete data, discrete data with more than two choices and continuous data, as discussed at length in O'Donoghue et al. (2020) .

For those with capital income, we assign the probability of holding shares across the age-income distribution on the basis of Monte Carlo estimates using Iterative Proportional Fitting (IPF) and we simulate an average change in the capital value or capital loss at the median (see Table C1 in Appendix C).

For work-related expenditures, we model and simulate commuting costs and childcare costs. For commuting costs, we first estimate the probability of commuting by car or by public transport as a function of occupation, industry; education, location, and age group (see Table A1 in Appendix A). Second, estimating models for both public transport and motor fuels as a function of household characteristics, disposable income, social group and number of workers, we predicted the proportional increase in these costs as a result of the number of workers in a household relative to not working. Without modelling the commuting distance as a function of income, which may have either a positive or a negative relationship, we assume a flat commuting cost across households, adjusted for the age.

The distribution of childcare costs per week by family type and disposable income decile is approximated using IPF. These averages are, in turn, used to calibrate the simulations based on the estimated models for having childcare and level of childcare expenditure (integrated in IGM, see Table B1 in Appendix B).

This allows us to "update" the microdata to the current period. We then use the infrastructure by introducing various shocks (e.g. factoring sector specific impacts, differentiated by age, macro changes, fiscal responses) and create counterfactual distributions.

The simulations involve two steps.
- First, we nowcast the lastest available survey data to December 2019 (assuming no COVID-19 crisis): $D(W_{t+1})$.
- Second, we assess the impact of COVID on the base 2020 income distribution by comparing the counterfactual distribution $D^*(W_{t+1}(L^*))$ under alternative shock scenarios (corresponding to different waves) to the "original" nowcasted distribution:
$$D(W_{t+1}) - D^*(W_{t+1}(L^*)).$$

## 3. Data and simulation assumptions

*Data*

As the main micro data source we use the 2017 version of the Survey on Income and Living Conditions (SILC). The SILC is a dataset that has been collected in Ireland since 2003 and which is used to form the Irish component of the European Union Statistics on Income and Living Conditions (EU-SILC). This representative survey contains information on socio-economic characteristics and incomes of households and living in them individuals, which is used for the construction of poverty and inequality indicators at the level of European Union. The Irish component is based on the information coming from two sources: a survey and register data. Overall, 80% of the respondents granted the permission to use their national social security numbers in order to access information on the benefit entitlement from administrative data (Callan et al., 2010).

In the context of this paper, the main advantage of the SILC data contains a rich set of variables needed for tax-benefit modelling. On the negative side, the dataset has a number of limitations, which might be challenging for microsimulation modelling. These include time mismatch in the measurement of income and personal characteristics, lack of information on some income components (e.g. wealth or property values) or tax-deductible expenditures (e.g. medical insurance), difficulties with attribution of some income variables to the appropriate unit of analysis (capital income, rental income, private transfers are recorded at the household level although they are often received by individuals), and aggregation of benefits. All these limitations are discussed in detail in O'Donoghue et al. (2013), whose strategy to address them we also follow in this paper.

Given that the 2017 SILC data contains income information, which refers to 2016, it cannot be used directly for the evaluation of the distributional impacts of the COVID-19 crisis. In order to account for the changes that elapsed between 2016 and 2020, we adjust the SILC data using a set of calibration control totals capturing the change in macroeconomic situation in Ireland over this period. This information is drawn from the Live-Register data and official statistics provided by the Irish Central Statistics Office. In what follows, we describe in more detail all the adjustments made to SILC data in order to make it timely appropriate for the analysis of the COVID-19 impacts as well as policy measures introduced to cushion individual incomes.

*Employment rate and sectoral impact*

Individuals who lose their job as a result of the COVID-19 crisis are eligible for a COVID-19 Pandemic Unemployment Payment. This instrument is available to workers who have lost their job on (or after) March 13. The payment is a flat rate non-means tested benefit, without additional payments for dependents paid to those aged 18-66.

The payment structure has changed a number of times over the crisis. In the first incarnation, at the start of the crisis on March 13 2020, it was at €203. This was increased in response to social partnership negotiations in relation to matters such as the levying of childcare fees to €350 per week on March 24 2020.

From June 29$^{th}$, payments were partially linked to previous earnings, with the introduction of several rates of payment: €300 per week for those with previous earnings of €300 or more, and €300, €250 and €203 for those with earning falling in the ranges €300-€400, €200-€300 and less than €200 respectively.

From September 17, there were two rates of payment: €203 for those earning less than €200 and €350 for all others.

From October 16 2020, there were three rates of payment: €350 per week for those with previous earnings of €400 or more, and €250 and €203 respectively for those earning in the ranges €200-€300 and less than €200.

From February 2021, there will be a reversion to two rates: €250 if earnings are over €300 and €203 otherwise.

The numbers and type of individuals eligible for payment and directly affected by the crisis are simulated using the income generation model. The overall employment rate is first used to calibrate the income generation model. This is characterised by the number of people in work relative to the population of a particular age group. The Labour Force Survey (LFS) collects data on this topic. However, as a quarterly survey, even with a relatively quick turn-around time from collection to publication, there is typically a 2-3 month lag between data collection and publication. In real time modelling within a period of economic volatility such as the COVID-19 crisis data that is closer to the period of the crisis is required.

The impact of the crisis is not a general demand shock, but a highly asymmetric change in employment, with "essential" industries remaining at work and some sectors such as the public sector remaining on full pay, while other industries are experiencing almost a full shut down over the period of the virus. There is relatively limited data in real time as to the sectoral impact of the crisis. The most suitable data to perform such calculations is the Administrative Data that is available on a monthly basis, typically 2-3 days after the end of the month, together with weekly updates in relation to aggregates that have been made. As is well documented, Live-Register data does not capture the level of unemployment equivalent to ILO measures. People can be working part-time whilst in receipt of benefits and conversely, someone can be out of work and seeking work, but not eligible for unemployment benefits. However, as an indicator in the short term, of a change in economic circumstances, the changes observed in the live register are an approximate indicator of changes in the numbers out of work (or non-employment rate). In this paper, the LFS is used to nowcast to December 2019, with the Administrative Data used to nowcast to May, June, August and November 2020.

Taking the change in the receipt of the pandemic unemployment benefit at the end of March 2020, we model the change in the employment rate at these 4 different points, reflecting both changes in the design of the instrument and reflecting different stages of the crisis. Table 1 outlines the assumed change in employment by sector, consistent with the overall change in live-register numbers as a result of COVID-19. We apply age specific changes identified in the Live Register and expressed as a proportion of the population in the SILC.

*COVID-19 Cases*

Individuals who have to stop working due to the COVID-19 infection or due to having been in a close contact with someone who contracted the disease are eligible for the COVID enhanced Illness Benefit (CEIB). The benefit is paid at the same rate as the PUP.

Both workers and non-workers get sick as a result of COVID-19. Table 2 outlines our random allocation of cases across in-work and out-of-work, within the national age distribution of the COVID-19 cases. Dividing by the proportion of workers in each age group, we derive the recipient rate of the COVID-19 related illness benefit.

**Table 1.** Pandemic Unemployment Payment and COVID Enhanced Illness Benefit changes over the first 9 months of the COVID-19 crisis

| Before Crisis | PUP | | | | CEIB | | | |
|---|---|---|---|---|---|---|---|---|
| | May 5th | June 6th | August 28th | November 15th | May 5th | June 6th | August 28th | November 15th |
| Agriculture, Forestry and Fishing; Mining and Quarrying | 8600 | 7100 | 3,200 | 4321 | 400 | 500 | 500 | 4321 |
| Manufacturing | 37400 | 28100 | 12,500 | 15465 | 5100 | 6600 | 7800 | 15465 |
| Electricity, gas supply; Water supply, sewerage and waste management | 2100 | 1700 | 900 | 1134 | 200 | 200 | 300 | 1134 |
| Construction | 79300 | 51500 | 17,500 | 21061 | 1700 | 1900 | 2200 | 21061 |
| Wholesale and Retail Trade; Repair of Motor Vehicles and motorcycles | 90300 | 76900 | 32,900 | 57015 | 8400 | 10400 | 11900 | 57015 |
| Transportation and storage | 17900 | 15400 | 9,600 | 9127 | 1600 | 1900 | 2200 | 9127 |
| Accommodation and food service activities | 128500 | 120000 | 48,700 | 102682 | 1600 | 1800 | 2400 | 102682 |
| Information and communication activities | 11800 | 11600 | 6,800 | 7526 | 700 | 800 | 900 | 7526 |
| Financial and insurance activities | 12500 | 11600 | 6,000 | 7119 | 1700 | 2100 | 2300 | 7119 |
| Real Estate activities | 8100 | 7600 | 3,500 | 5442 | 300 | 400 | 500 | 5442 |
| Professional, Scientific and Technical activities | 24800 | 22300 | 11,700 | 13294 | 1300 | 1600 | 1800 | 13294 |
| Administrative and support service activities | 45800 | 41100 | 23,600 | 29674 | 3900 | 4900 | 5500 | 29674 |
| Public Administration And Defence; Compulsory Social Security | 14400 | 11700 | 5,600 | 5354 | 1700 | 2000 | 2200 | 5354 |
| Education | 22000 | 21600 | 14,400 | 10340 | 600 | 800 | 900 | 10340 |
| Human Health And Social Work activities | 22500 | 19700 | 9,900 | 10271 | 8300 | 10900 | 12200 | 10271 |
| Arts, entertainment and recreation | 14200 | 13800 | 6,300 | 11973 | 200 | 300 | 400 | 11973 |
| Other Sectors | 39200 | 37500 | 10,300 | 31048 | 1200 | 1400 | 1700 | 31048 |

Note: Employment is expressed in number of individuals.

**Table 2. Distribution of COVID-19 cases by age group and by work status (April 2020)**

| | Age Group | | | | | | | | |
|---|---|---|---|---|---|---|---|---|---|
| | 0 | 1-4 | 5-14 | 15-24 | 25-34 | 35-44 | 45-54 | 55-64 | 65+ |
| In-work by Age | 0 | 0 | 0 | 91 | 413 | 452 | 441 | 259 | 61 |
| Out-of-Work by Age | 9 | 12 | 33 | 164 | 265 | 299 | 323 | 325 | 857 |

Source: COVID-19 Dashboard
Notes: the same approach was utilised to simulate COVID-19 cases for other age groups (https://geohive.maps.arcgis.com/apps/opsdashboard/index.html#/29dc1fec79164c179d18d8e53df82e96), accessed April 6th 2020.

*Mortgage Interest*

Individuals who have to endure mortgage repayments received a possibility to freeze up to 3 months of such repayments during the COVID-19 crisis. This resulted in 28000 applications for mortgage deferrals in March 2020 with the numbers going steadily up in subsequent months. As of July, 160000 deferrals had been made (see Table 3).

**Table 3. Number of requests for mortgage deferral**

| | |
|---|---|
| Number of Requests as of March 28 | 28000 |
| Number of Requests as of April 12 | 45000 |
| Number of Requests as of July | 160000 |
| Number of Requests as of September | 90539 |

Source: https://www.rte.ie/news/business/2020/0328/1127000-banking-mortgages-coronavirus/
https://www.irishexaminer.com/breakingnews/ireland/mortgage-breaks-for-six-months-as-45000-apply-for-payment-pause-993714.html
https://www.centralbank.ie/statistics/statistical-publications/behind-the-data/covid-19-payment-breaks-who-has-needed-them

*Work Related Expenditures*

Due to intensified work from home and business closures, the size of work-related expenses, such as commuting costs and childcare costs, decreased substantially for most workers. In order to account for the reductions in these expenses we use the Household Budget Survey data from 2016 to simulate the amount of commuting costs typically incurred by an average worker (Table 4). It should be noted that those who do not work also have transport costs for other purposes. While the actual cost of commuting for work may be higher, it is assumed that there would be some substitution if an individual was not working.

**Table 4. Cost of commuting per week**

| | Number of Workers | | |
|---|---|---|---|
| | 1 | 2 | 3 |
| Proportional Increase in Cost relative to not working | | | |
| Motor Fuels | 0.263 | 0.482 | 0.721 |
| Public Transport | 0.172 | 0.253 | 0.595 |
| Cost per week | | | |
| Motor Fuels | 7.41 | 13.59 | 20.33 |
| Public Transport | 1.76 | 0.83 | 3.49 |
| Total per week (€) | 9.17 | 14.42 | 23.82 |

Source: Household Budget Survey 2015-16.

Following the outbreak of the pandemics, the State took a decision to support childcare providers in order to maintain the sustainability of the childcare sector and relieve parents from childcare payments while keeping childcare places. [6] Utilising IPF to data collected within the Household Budget Survey, we derive the distribution of childcare costs per week by family type and disposable income decile (Table 5). These averages are simulated across households in the sample on the basis of the regressions outlined in Table C.1 in Appendix C.

Table 5. Distribution of Child Care Costs per Week by Family Type and Disposable Income Decile

| Family Type | 1 | 2 | 3 | 4 | 5 | 6 | 7 | 8 | 9 | 10 | Total |
|---|---|---|---|---|---|---|---|---|---|---|---|
| 1 adult with children | 2.9 | 7.8 | 3.3 | 22.0 | 22.4 | 39.1 | 68.0 | 65.9 | 191.2 | 268.5 | 18.1 |
| 2 adults with 1-3 children | 1.9 | 5.1 | 2.2 | 14.5 | 14.7 | 25.8 | 44.8 | 43.4 | 126.1 | 177.0 | 49.9 |
| Other households with children | 0.7 | 2.0 | 0.8 | 5.6 | 5.7 | 9.9 | 17.2 | 16.7 | 48.4 | 68.0 | 15.2 |
| Total | 0.4 | 1.0 | 0.6 | 5.5 | 4.7 | 7.6 | 12.9 | 13.6 | 30.2 | 40.3 | 12.0 |

Source: Household Budget Survey 2015-16

It is assumed in the simulations that those who receive Pandemic Payments or those who are non-essential workers working from home do not incur commuting costs or childcare expenses.

*Pandemic Wage Subsidy*

On March 24 2020 the State introduced a COVID-19 Temporary Wage Subsidy Scheme to ensure that employers would keep their employees even if the revenues go down. In order to be eligible for the scheme businesses would need to have a minimum of 25% decline in turnover. For the initial period between March 13 and March 24, the scheme was a flat rate payment of €203. From March 26 to April 20, it was redesigned to cover up to 70% of an employee's net earnings. The payment was limited, however, depending on the employee's average take home pay:
- Average pay from €0 to €586 limits it to €410;
- Average pay from €586 to €960 limits it to €350;
- Average pay above €960 is not entitled to the subsidy.

On April 20[th], the rates of temporary wage subsidy were changed as follows:
- 70% to 85% for employees with a previous average take home pay below €412 per week
- €350 per week for employees with a previous average take home pay between €412 and €500 per week
- The subsidy remained the same for employees with a previous average take home pay of between €500 and €586 per week
- A tiered system has been introduced for employees with a previous average take home pay of over €586 per week
- Employees who were taking home more than €960 per week would be able to avail of the scheme, with tapers depending upon the proportion paid by the employer.

From the 1[st] of September 2021, the Temporary Wage Subsidy Scheme was replaced by the Employment Wage Subsidy Scheme, where employers and new firms in sectors impacted by COVID-19 whose turnover has fallen 30% get a flat-rate subsidy per week.

---

6 https://www.gov.ie/en/press-release/e37415-minister-katherine-zappone-announces-measures-to-support-childcare-p/

Between 1 July 2020 and 19 October 2020, the following subsidy rates applied:

| Gross pay per week | Revised subsidy rates |
|---|---|
| Less than €151.50 | No subsidy applies |
| €151.50 - €202.99 | €151.50 |
| €203 - €1,462 | €203 |
| Over €1,462 | No subsidy applies |

The subsidy rates from 20 October 2020 to 31 January 2021 are

| Gross pay per week | Revised subsidy rates |
|---|---|
| Less than €151.50 | No subsidy applies |
| €151.50 - €202.99 | €203 |
| €203 - €299.99 | €250 |
| €300 - €399.99 | €300 |
| €400 - €1,462 | €350 |
| Over €1,462 | No subsidy applies |

The Wage Subsidy itself has a limited distributional impact but it shifts the burden of payments from the private sector to the public sector. This subsidy also does not take into account the impact of wage reductions where employers did not have the cash flow to make these payments as in the case of individuals whose take-home pay exceeds the wage subsidy limit. Prior to the introduction of the subsidy scheme there had been pay reductions for staff in certain sectors most affected by the crisis, where staff were not made redundant, such as the airline sector. For example, the two main airlines (Ryanair and Aer Lingus) halved the pay of staff when flights were grounded.[7]

*Stock Market*

The outbreak of the COVID-19 pandemics lead to a fall in stock markets. Only between January 1 and April 1 2020 the Irish index ISEQ fell by 32% which had a further impact on the financial situation of households holding shares. Using data from the Household Finance and Consumption Survey in 2018, Table 6 reports how the holding of shares is spread along the distributions of household income and age. It shows that individuals located in the top quantile of the household income distribution are 8 times more likely to have financial assets than those located in the bottom quantile of the distribution, with the values of assets being 9 times higher. The distribution of financial assets across the age distribution is not as extreme, with those aged 40-79 more likely than other age groups to hold shares.

The data equivalent to Table 7, with change in share values by age and income group, was not available during the analysis of this paper. In order to utilise this information in a microsimulation model, Iterative Proportional Fitting (IPF) was used to create an approximation of the share value holdings across the age- income distribution. The average share holding and the median value of holdings were generated separately and then multiplied to get the average value per person in the cell (see Appendix D). Applying the ISEQ index to January 1 2020 and then to April 1, 2020, Table 7 models the net change in the value of shares across the age-income distribution. The Table shows that the biggest losses were experienced by those with the highest incomes and the oldest.

---

[7] https://www.irishtimes.com/business/work/coronavirus-employers-should-seek-consent-for-pay-cuts-lawyer-1.4221405.

Table 6. Distribution of holding and value of shares, 2018

| Percentile of household income | Less than 20 | 20-39 | 40-59 | 60-79 | 80-100 |
|---|---|---|---|---|---|
| Participation in total financial assets (%) | 3.3 | 3 | 8.3 | 11.3 | 24.8 |
| Median values of financial assets (€Thousand) | 1.4 | 8.8 | 3.1 | 4.4 | 12.2 |
| Distribution of total financial assets (%) | 1.4 | 4.3 | 11 | 12.5 | 12.4 |
| | | | | | |
| Age group | Under 35 years | 35 - 44 years | 45 - 54 years | 55 - 64 years | 65 years and over |
| Participation in total financial assets (%) | 5.4 | 8.7 | 13.3 | 13.8 | 8.3 |
| Median values of financial assets (€Thousand) | 14.1 | 8.4 | 4 | 10 | 12.9 |
| Distribution of total financial assets (%) | 4.6 | 15.3 | 11.9 | 5.4 | 15.5 |

Source: Household Finance and Consumption Survey.

Table 7. Change in shareholdings across the age-income distribution, January 1 – April 1, 2020 (€000)

| Age group | Percentile in the income distribution | | | | | |
|---|---|---|---|---|---|---|
| | Less than 20 | 20-39 | 40-59 | 60-79 | 80-100 | Total |
| 30 | 0.000 | -0.004 | -0.003 | -0.006 | -0.011 | -0.005 |
| 40 | -0.002 | -0.036 | -0.032 | -0.063 | -0.117 | -0.055 |
| 50 | -0.003 | -0.047 | -0.041 | -0.082 | -0.151 | -0.072 |
| 60 | -0.012 | -0.194 | -0.168 | -0.336 | -0.623 | -0.246 |
| 70 | -0.058 | -0.902 | -0.783 | -1.563 | -2.901 | -0.698 |
| Total | -0.025 | -0.248 | -0.134 | -0.197 | -0.328 | -0.183 |

Note: a similar approach was applied to later periods.

## 4. Results

*Average and Distributional Analysis*

Table 8 reports the trend in average equivalised incomes using three definitions:
- Market Income
- Disposable Income
- Adjusted Disposable Income

Market income, unsurprisingly, fell over the course of the crisis. The decline was the most drastic in the first wave (around 32 percent) with a slight recovery taking place in June and August. However, the onset of the second wave in November saw a reduction in market incomes again, albeit to a much lower degree than in May, reflecting the fact that the restrictions were not as great. The end of the second wave at Christmas, saw a slight improvement, albeit not to the same degree as in the summer, before worsening again after Christmas in the 3$^{rd}$ wave. However, sectors affected in the second wave were more likely to be concentrated at the bottom of the distribution than at the top in the second and third wave, rather than the first wave.

The drop in market incomes was partially compensated by the increase in benefit payments, which can be seen from the trends in disposable incomes. The average size of disposable income decreased by almost 10 percent at the start of the crisis, before making a slight recovery later on. Nevertheless, it remained 6-7% below the pre-crisis level during the second and third waves of the pandemic.

As mentioned above, the onset of the COVID-19 crisis has pushed a substantial share of employees to work from home or to take up temporary unemployment. This led to a decrease in commuting costs and childcare expenses. In addition, some individuals applied for mortgage deferrals, which further reduced their current expenditures. All these expenses are captured with the adjusted disposable income, which remained largely unchanged over the course of the crisis.

**Table 8. Average income by income definition over the course of the crisis**

|  | Before Crisis | May 5th | June 6th | August 28th | November 13th | December 22nd | January 26th |
|---|---|---|---|---|---|---|---|
| Markety Income | 2362 | 1600 | 1675 | 1931 | 1810 | 1791 | 1817 |
| Disposable Income | 2150 | 1945 | 1957 | 2020 | 2005 | 2003 | 2000 |
| Disposable Income* | 1861 | 1798 | 1810 | 1868 | 1853 | 1853 | 1853 |

Hidden below these averages are quite a differential impact on different parts of the income distribution. Figure 3 summarizes the distribution of market (Panel A), disposable (Panel B) and adjusted disposable (Panel C) incomes before the crisis and at various points during the crisis calculated per adult equivalent. Deciles are calculated based upon adjusted disposable income decile, adjusted for work, housing expenses and capital losses.

In terms of market income, the decline was larger at the top than at the bottom of the distribution, however proportionally the loss at the top was slightly less than the bottom. Disposable incomes at the bottom of the distribution were largely maintained over the three waves and even higher in some cases where benefits were higher than pre-crisis market income. The gradual targeting of payments in later waves reduced the incidence of those at the bottom of the income distribution with disposable incomes that were higher than at the pre-crisis level. Disposable incomes at the top of the distribution fell by 10-20% at the start of the crisis, and recovered to about 10% in later waves.

Panel C in Figure 3 suggests that adjusted household equivalized disposable incomes were somewhat cushioned during the crisis, which held them from the same decline as we observe in market incomes. Furthermore, there was even a rise in adjusted equivalised disposable incomes amongst lower deciles in the first and second waves. This reflects both the generosity of the instruments and the fact that the squeezed middle, who are at the bottom of the distribution have normally both low incomes and high fixed costs of working and housing. This progressivity was visible for all periods. However, we see that the differential proportional change between the top and the bottom of the distribution has shrunk in the second and third wave as a result of the increased targeting of benefits that occurred later on.

**Figure 3. Distributional characteristics of income before and during crisis (€ per month per adult equivalent)**

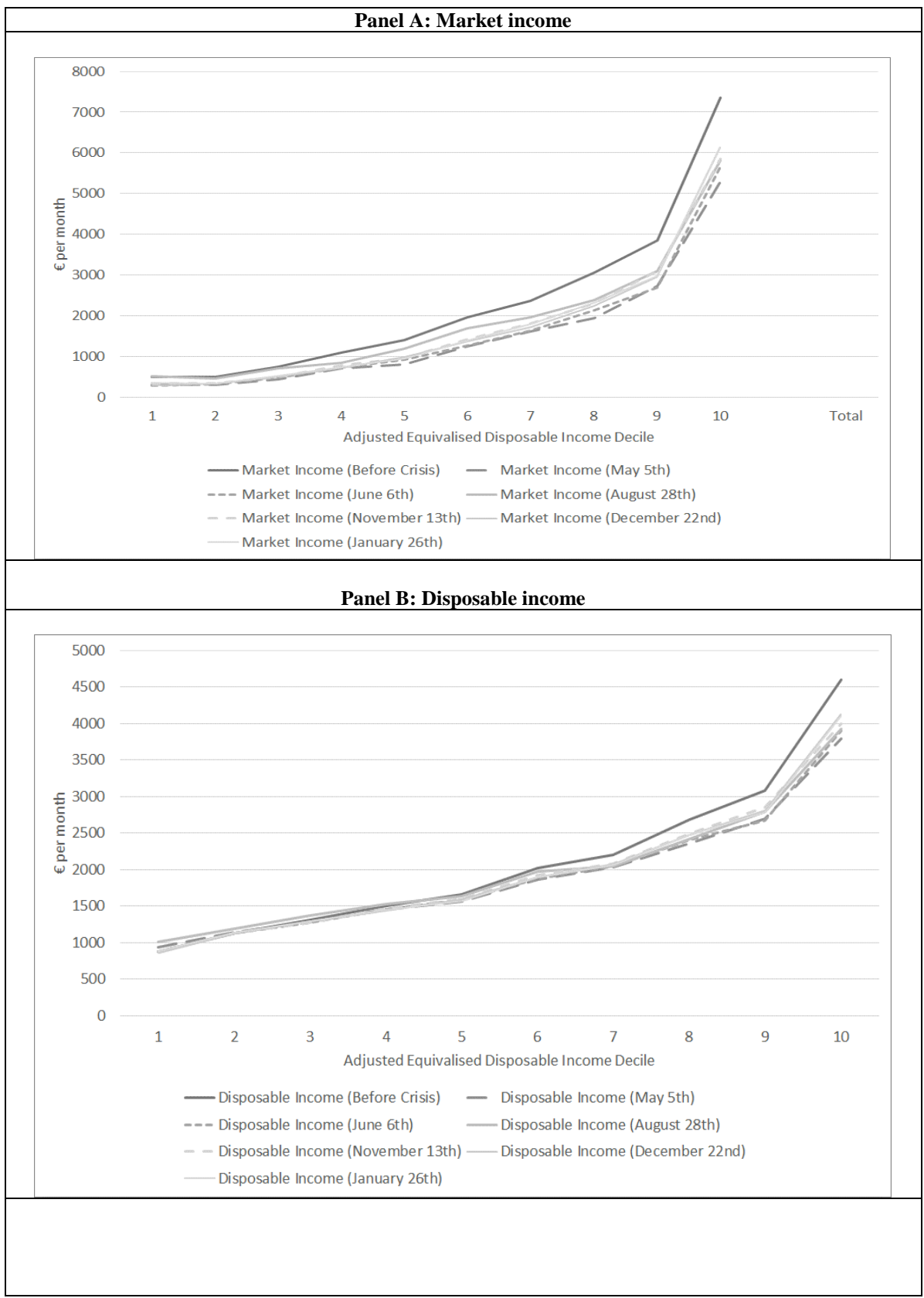

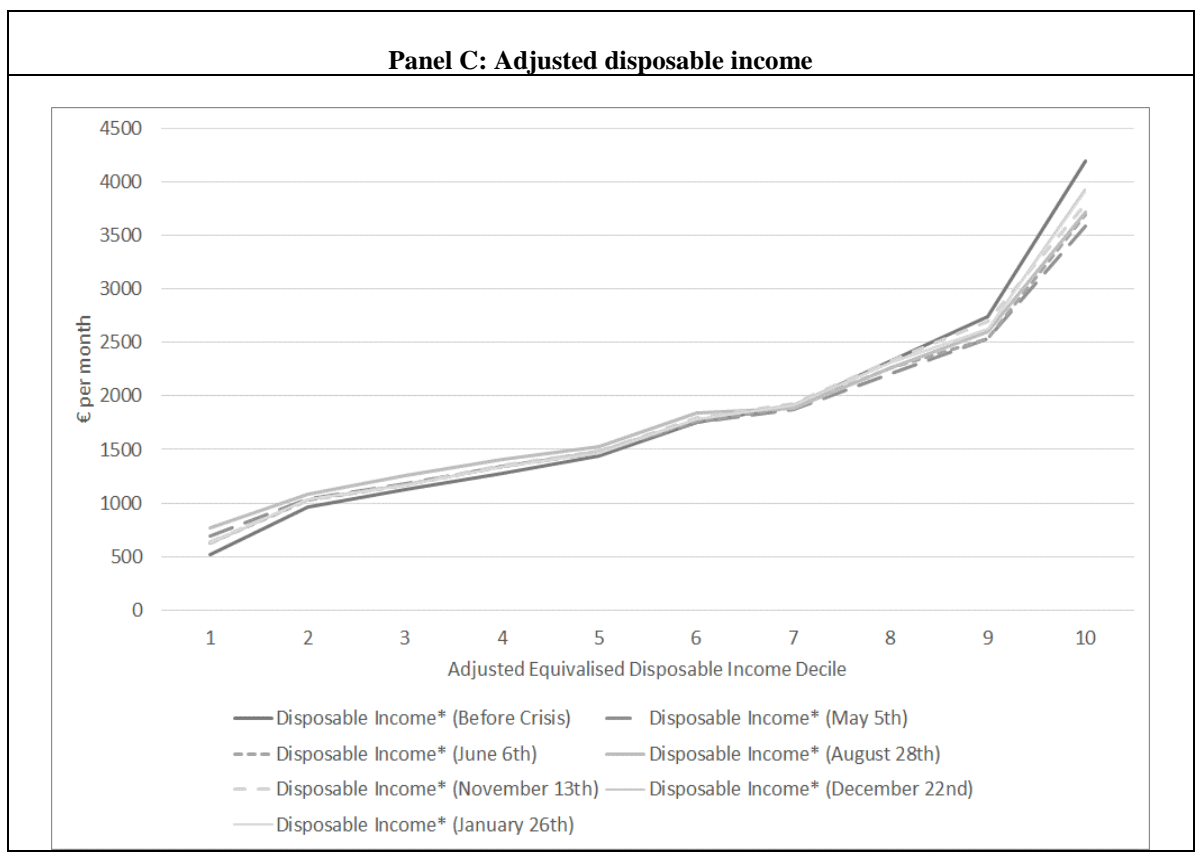

Note: Adjusted disposable income stands for household equivalized disposable income adjusted for housing, work related expenses and capital losses.

*Inequality and Redistribution*

Table 9 summarizes changes in inequality of different types of incomes during six time points over the crisis as compared to the pre-crisis period. It shows that inequality in market income increased by 0.12 points at the start of the crisis, declining gradually as the economy reopened and then increased again as the country went into the second and third waves. In contrast, inequality in gross income, disposable income, and especially adjusted disposable income decreased throughout the crisis. The largest declines were observed at the pick of the first and second waves following the strengthening of the lockdown measures, activation of COVID-19 related benefit payments, and decreases in work-related expenses.

Table 10 reports the redistributive impact of public policy. The contribution of benefits to redistribution is derived as the difference in the Gini coefficients calculated for gross and market incomes. The contribution of taxes to redistribution is derived as the difference in the Gini coefficients calculated for disposable and gross incomes. The contribution of work-related and housing costs to redistribution is derived as the difference in the Gini coefficients for disposable income adjusted for work-related and housing expenditures and disposable income without these adjustments. Out of three policy instruments, the redistributive role of benefits increased the most over the course of the crisis. It was the highest at the pick of the first wave and slightly declined afterwards remaining, nevertheless, much higher than in the pre-crisis period. In contrast, the redistributive role of taxes and work-related expenses remained relatively stable during the first three waves of the COVID-19 crisis. In general, taxes help to decrease inequality in incomes whereas the opposite applies to work-related expenses.

Table 9. Gini coefficient before and during crisis

|  | Market Income | Gross Income | Disposable Income | Disposable Income* |
|---|---|---|---|---|
| Gini |  |  |  |  |
| Before Crisis | 0.490 | 0.363 | 0.290 | 0.308 |
| May 5th | 0.609 | 0.349 | 0.276 | 0.290 |
| June 6th | 0.594 | 0.354 | 0.279 | 0.294 |
| August 28th | 0.548 | 0.361 | 0.291 | 0.304 |
| November 15th | 0.572 | 0.356 | 0.282 | 0.296 |
| December 22nd | 0.582 | 0.362 | 0.287 | 0.301 |
| January 26th | 0.578 | 0.361 | 0.287 | 0.301 |
| Change |  |  |  |  |
| May 5th | 0.119 | -0.015 | -0.015 | -0.019 |
| June 6th | 0.104 | -0.009 | -0.011 | -0.014 |
| August 28th | 0.058 | -0.002 | 0.001 | -0.004 |
| November 15th | 0.082 | -0.007 | -0.009 | -0.012 |
| December 22nd | 0.092 | -0.001 | -0.003 | -0.007 |
| January 26th | 0.088 | -0.002 | -0.004 | -0.008 |

Note: Disposable income* stands for household equivalized disposable income adjusted for housing, work related expenses and capital losses. The Modified OECD Equivalence Scale is used.

Table 10. Redistribution before and during crisis

| Redistribution | Benefits | Taxes | Work Expenses and Housing Costs |
|---|---|---|---|
| Before Crisis | -0.127 | -0.073 | 0.018 |
| May 5th | -0.260 | -0.073 | 0.014 |
| June 6th | -0.240 | -0.075 | 0.016 |
| August 28th | -0.187 | -0.070 | 0.014 |
| November 15th | -0.216 | -0.074 | 0.014 |
| December 22nd | -0.220 | -0.075 | 0.014 |
| January 26th | -0.217 | -0.074 | 0.014 |

## 5. Discussion and conclusions

This paper uses a microsimulation framework to undertake an analysis of the distributional implications of the COVID crisis. Given the lack of real-time data during the fast moving crisis, it applies a nowcasting methodology in combination with real-time aggregate administrative data to calibrate an income survey and to simulate changes in the tax benefit system that attempted to mitigate the impacts of the crisis. The paper builds upon earlier work that was undertaken at an early phase of the crisis and assesses the impact over three waves of the COVID-19 crisis. It also describes in detail the methodology used to derive an adjusted disposable income measures, accounting for the impact of non-discretionary expenditures.

Our analysis shows that, despite the increase in market income inequality, inequality in disposable incomes decreased, mainly due to the combination of crisis-induced discretionary policy measures with existing system of taxes and benefits. The decrease in disposable income inequality was the largest during the first wave of the COVID-19 pandemic but became somewhat smaller afterwards following a reduction in the generosity of benefits as they became more targeted.

The paper also demonstrates how an approach that combines microsimulation and nowcasting can provide policy makers with real-time data needed for elaboration of timely policy interventions as crisis unfolds in the situation when survey data comes with time lags. Needless to say, the application of this methodology requires a number of assumptions. However, with careful sensitivity analysis, the model provides a flexible tool to policy designers to explore the implications of alternative assumptions in addition to alternative policies.

On a methodological level, our paper makes a specific contribution in relation to the choice of welfare measure in assessing the impact of the COVID-19 crisis on inequality. As the impact is multi-faceted involving market income, public policy and changes in fixed costs of work and housing, we adopt a novel measure adjusting equivalised household disposable income to account for changes in housing, child care and commuting costs and reflects impacts of changes in capital values. The modelling approach illustrated in this paper can inform the trade-offs between measures – e.g. attenuating the rise in inequality or poverty while limiting the fiscal costs – that are inherent in the policy-design challenge.

**Appendix A: Phases of the COVID-19 crisis and total expenditure in Ireland**

**Figure A.1. Phases of the COVID-19 crisis and total expenditure in Ireland, per user (indexed to 1stFebruary, 7-day moving average)**

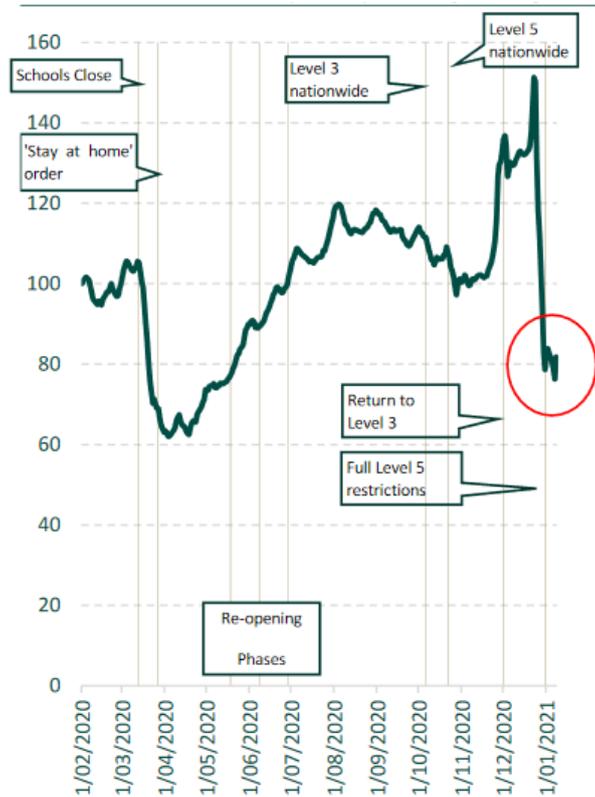

Source: Revolut in Department of Finance Emerging economic developments-real-time economic domesticindicators. 11November 2020.

# Appendix B: Estimation of probabilities of using various types of transport

## Table B.1 Probability of using public transport or private transport

|  | Public Transport | | | Private Transport \| Not Public Transport | | |
|---|---|---|---|---|---|---|
|  | coef | S.E. | p-vaue | coef | S.E. | p-vaue |
| Manufacturing industries, mining, quarrying and turf production, electricity, gas and water supply | 0.692 | 0.073 | 0.000 | 0.677 | 0.022 | 0.000 |
| Construction | 0.362 | 0.076 | 0.000 | 1.214 | 0.024 | 0.000 |
| Commerce | 1.314 | 0.072 | 0.000 | 0.145 | 0.021 | 0.000 |
| Transport Storage Communications | 2.179 | 0.072 | 0.000 | 0.138 | 0.021 | 0.000 |
| Public administration and defence | 1.719 | 0.073 | 0.000 | 0.846 | 0.023 | 0.000 |
| Education, health and social work | 1.167 | 0.073 | 0.000 | 0.532 | 0.021 | 0.000 |
| Other | 1.424 | 0.073 | 0.000 | 0.043 | 0.022 | 0.055 |
| Border Midland and Wester Region | -1.457 | 0.011 | 0.000 | 0.257 | 0.005 | 0.000 |
| Occupation 1 | 0.148 | 0.013 | 0.000 | 0.697 | 0.008 | 0.000 |
| Occupation 2 | 0.098 | 0.015 | 0.000 | 0.412 | 0.010 | 0.000 |
| Occupation 3 | 0.044 | 0.014 | 0.002 | 0.559 | 0.009 | 0.000 |
| Occupation 4 | 0.402 | 0.012 | 0.000 | 0.287 | 0.007 | 0.000 |
| Occupation 5 | -1.643 | 0.099 | 0.000 | -1.369 | 0.023 | 0.000 |
| Occupation 6 | -0.918 | 0.104 | 0.000 | 0.156 | 0.028 | 0.000 |
| Occupation 7 | -0.342 | 0.018 | 0.000 | 0.926 | 0.010 | 0.000 |
| Occupation 8 | 0.040 | 0.015 | 0.009 | 0.259 | 0.009 | 0.000 |
| Aged 20-24 | -0.439 | 0.028 | 0.000 | 0.802 | 0.022 | 0.000 |
| Aged 25-29 | -0.705 | 0.027 | 0.000 | 1.136 | 0.022 | 0.000 |
| Aged 30-34 | -0.906 | 0.027 | 0.000 | 1.467 | 0.022 | 0.000 |
| Aged 35-39 | -1.147 | 0.027 | 0.000 | 1.662 | 0.022 | 0.000 |
| Aged 40-44 | -1.322 | 0.028 | 0.000 | 1.677 | 0.022 | 0.000 |
| Aged 45-49 | -1.377 | 0.028 | 0.000 | 1.613 | 0.022 | 0.000 |
| Aged 50-54 | -1.334 | 0.028 | 0.000 | 1.489 | 0.022 | 0.000 |
| Aged 55-59 | -1.289 | 0.029 | 0.000 | 1.363 | 0.022 | 0.000 |
| Aged 60-64 | -1.287 | 0.031 | 0.000 | 1.172 | 0.023 | 0.000 |
| Aged 65-69 | -1.350 | 0.041 | 0.000 | 0.816 | 0.026 | 0.000 |
| Aged 70-74 | -1.471 | 0.066 | 0.000 | 0.431 | 0.033 | 0.000 |
| Aged 75+ | -1.606 | 0.091 | 0.000 | -0.035 | 0.039 | 0.360 |
| University Education | 0.242 | 0.007 | 0.000 | 0.016 | 0.005 | 0.002 |
| Constant | -2.839 | 0.077 | 0.000 | -0.988 | 0.030 | 0.000 |
| Pseudo R2 | 0.109 | | | 0.089 | | |
| Number of Obs | 1682588 | | | 1682588 | | |

Note : Calculated on the basis of Census of Population Data.

**Appendix C: Simulation of the child care participation and costs**

**Table C.1. Regression models for having child care (Logit) and level of childcare expenditure**

|  | Coef. | Std. Err. | P>z | Coef. | Std. Err. | P>z |
|---|---|---|---|---|---|---|
|  | Has Child Care | | | Child Care Expenditure | | |
| Number of Children Aged 0 -4 | 0.833 | 0.073 | 0.000 | 28.0 | 4.8 | 0 |
| Number of Children | -0.018 | 0.057 | 0.750 | 0.0 | 4.0 | 0.992 |
| Disposable Income (Equivalised) | 0.003 | 0.000 | 0.000 | 0.1 | 0.0 | 0 |
| Disposable Income (Equivalised) Squared | 0.000 | 0.000 | 0.000 | | | |
| Number of Workers = 2 \| Lone Parent Working | 1.224 | 0.129 | 0.000 | 54.0 | 9.6 | 0 |
| Constant | -3.584 | 0.246 | 0.000 | -15.5 | 13.1 | 0.238 |
| R2 | | | | 0.1437 | | |
| Pseudo R2 | 0.1836 | | | | | |
| Observations | 1,937 | | | 719 | | |

Note : Calculated on the basis of Household Budget Survey 2015-2016.

**Appendix D**

Approximation of the share value holdings across the age-income distribution

**Table D.1. Age-income distribution of shareholdings proportion, 2018**

| Age group | Percentile in the income distribution | | | | | |
|---|---|---|---|---|---|---|
|  | Less than 20 | 20-39 | 40-59 | 60-79 | 80-100 | Total |
| 30 | 0.012643 | 0.013099 | 0.039393 | 0.056208 | 0.12114 | 0.054 |
| 40 | 0.020834 | 0.021585 | 0.064915 | 0.092623 | 0.199625 | 0.087 |
| 50 | 0.030946 | 0.032062 | 0.096424 | 0.137582 | 0.296521 | 0.133 |
| 60 | 0.037694 | 0.039053 | 0.117449 | 0.167581 | 0.361176 | 0.138 |
| 70 | 0.037777 | 0.039139 | 0.117707 | 0.16795 | 0.36197 | 0.083 |
| Total | 0.029736 | 0.027031 | 0.074783 | 0.101811 | 0.223442 | 0.09066 |

Source: Household Finance and Consumption Survey, with Iterative Proportional Fitting

**Table D.2. Age-income distribution of shareholdings €000, 2018**

| Age group | Percentile in the income distribution | | | | | |
|---|---|---|---|---|---|---|
|  | Less than 20 | 20-39 | 40-59 | 60-79 | 80-100 | Total |
| 30 | 0.001 | 0.010 | 0.009 | 0.018 | 0.033 | 0.016 |
| 40 | 0.007 | 0.103 | 0.090 | 0.179 | 0.332 | 0.156 |
| 50 | 0.009 | 0.133 | 0.116 | 0.231 | 0.428 | 0.205 |
| 60 | 0.035 | 0.548 | 0.476 | 0.950 | 1.763 | 0.698 |
| 70 | 0.164 | 2.554 | 2.217 | 4.427 | 8.214 | 1.976 |
| Total | 0.069 | 0.703 | 0.379 | 0.558 | 0.930 | 0.518 |

Source: Household Finance and Consumption Survey, with Iterative Proportional Fitting

**Appendix E**

**Table E.1 Distributional characteristics of income before and during crisis (€ per month per adult equivalent)**

| Decile | Before Crisis | | | | May 5th | | | | June 6th | | | | August 28th | | | |
|---|---|---|---|---|---|---|---|---|---|---|---|---|---|---|---|---|
| | Market Income | Gross Income | Disposable Income | Disposable Income* | Market Income | Gross Income | Disposable Income | Disposable Income* | Market Income | Gross Income | Disposable Income | Disposable Income* | Market Income | Gross Income | Disposable Income | Disposable Income* |
| 1 | 504.0 | 960.3 | 878.6 | 522.2 | 306.5 | 1021.6 | 933.1 | 694.7 | 281.5 | 949.0 | 871.8 | 631.1 | 517.6 | 1135.2 | 1012.6 | 765.1 |
| 2 | 501.6 | 1294.6 | 1132.5 | 957.7 | 296.4 | 1299.7 | 1142.6 | 1043.8 | 316.7 | 1279.4 | 1124.9 | 1023.5 | 453.5 | 1360.5 | 1188.2 | 1081.8 |
| 3 | 746.0 | 1518.1 | 1315.7 | 1121.9 | 436.6 | 1480.5 | 1289.7 | 1179.8 | 477.1 | 1468.9 | 1278.1 | 1171.1 | 708.4 | 1630.9 | 1376.4 | 1260.7 |
| 4 | 1087.8 | 1765.3 | 1498.0 | 1278.3 | 706.7 | 1707.1 | 1455.7 | 1341.9 | 730.5 | 1704.4 | 1455.7 | 1339.2 | 839.0 | 1786.5 | 1524.5 | 1405.4 |
| 5 | 1398.3 | 2005.1 | 1665.3 | 1434.7 | 810.8 | 1842.5 | 1563.3 | 1471.1 | 912.7 | 1898.3 | 1590.0 | 1480.8 | 1196.8 | 1988.1 | 1636.2 | 1531.8 |
| 6 | 1959.9 | 2579.5 | 2020.2 | 1757.4 | 1258.4 | 2326.3 | 1861.1 | 1755.0 | 1272.9 | 2360.1 | 1880.4 | 1769.5 | 1689.9 | 2507.1 | 1972.0 | 1844.2 |
| 7 | 2361.9 | 2847.5 | 2203.0 | 1907.2 | 1613.6 | 2582.7 | 2028.6 | 1869.7 | 1642.3 | 2604.7 | 2044.0 | 1889.5 | 1966.8 | 2648.9 | 2046.3 | 1887.3 |
| 8 | 3053.2 | 3542.6 | 2680.3 | 2326.0 | 1946.1 | 3053.2 | 2354.0 | 2204.8 | 2139.6 | 3123.1 | 2400.6 | 2260.8 | 2377.3 | 3146.4 | 2423.9 | 2263.1 |
| 9 | 3859.7 | 4270.8 | 3083.2 | 2740.6 | 2740.6 | 3677.0 | 2694.9 | 2535.1 | 2694.9 | 3631.3 | 2672.1 | 2535.1 | 3106.0 | 3814.0 | 2786.3 | 2603.6 |
| 10 | 7350.1 | 7620.7 | 4599.5 | 4193.6 | 5298.4 | 6087.5 | 3787.8 | 3584.9 | 5659.1 | 6403.2 | 3900.5 | 3697.6 | 5817.0 | 6335.5 | 3923.1 | 3720.1 |
| Total | 2362.0 | 2916.3 | 2149.9 | 1860.7 | 1600.4 | 2554.8 | 1945.0 | 1798.0 | 1675.1 | 2603.0 | 1957.1 | 1810.0 | 1930.6 | 2699.4 | 2019.7 | 1867.9 |

| Decile | November 13th | | | | December 22nd | | | | January 26th | | | | |
|---|---|---|---|---|---|---|---|---|---|---|---|---|---|
| | Market Income | Gross Income | Disposable Income | Disposable Income* | Market Income | Gross Income | Disposable Income | Disposable Income* | Market Income | Gross Income | Disposable Income | Disposable Income* | |
| 1 | 342.8 | 962.6 | 880.9 | 635.7 | 295.1 | 933.1 | 862.7 | 624.3 | 347.4 | 958.1 | 878.6 | 635.7 | |
| 2 | 352.2 | 1289.6 | 1132.5 | 1031.1 | 334.4 | 1287.0 | 1129.9 | 1028.6 | 334.4 | 1289.6 | 1135.0 | 1031.1 | |
| 3 | 500.2 | 1463.2 | 1278.1 | 1171.1 | 508.9 | 1492.1 | 1295.4 | 1176.9 | 523.4 | 1477.6 | 1286.8 | 1168.2 | |
| 4 | 778.1 | 1725.6 | 1471.5 | 1352.4 | 719.9 | 1683.3 | 1442.4 | 1333.9 | 738.4 | 1696.5 | 1453.0 | 1341.9 | |
| 5 | 954.0 | 1903.2 | 1597.3 | 1488.1 | 990.4 | 1903.2 | 1590.0 | 1483.2 | 944.3 | 1869.2 | 1570.6 | 1468.7 | |
| 6 | 1432.0 | 2434.8 | 1918.9 | 1796.0 | 1374.1 | 2381.8 | 1890.0 | 1779.1 | 1403.0 | 2393.8 | 1897.2 | 1781.5 | |
| 7 | 1816.7 | 2670.9 | 2086.0 | 1929.3 | 1726.2 | 2648.9 | 2072.7 | 1918.2 | 1796.8 | 2648.9 | 2066.1 | 1909.4 | |
| 8 | 2302.7 | 3262.9 | 2493.8 | 2330.7 | 2232.8 | 3216.3 | 2470.5 | 2312.0 | 2319.0 | 3239.6 | 2470.5 | 2316.7 | |
| 9 | 3106.0 | 3905.4 | 2854.8 | 2694.9 | 2946.1 | 3836.8 | 2809.1 | 2626.4 | 2969.0 | 3791.2 | 2786.3 | 2626.4 | |
| 10 | 5862.0 | 6515.9 | 3990.7 | 3787.8 | 6132.6 | 6763.9 | 4126.0 | 3923.1 | 6132.6 | 6718.8 | 4103.4 | 3900.5 | |
| Total | 1810.0 | 2675.3 | 2005.3 | 1853.4 | 1790.8 | 2675.3 | 2002.9 | 1853.4 | 1817.3 | 2675.3 | 2000.4 | 1853.4 | |

Note: Disposable income* stands for household equivalized disposable income adjusted for housing, work related expenses and capital losses.